\documentclass[superscriptaddress,preprint]{revtex4-1}
\usepackage{amssymb, amsmath}
\usepackage{graphicx}
\newcommand{\fcs}[0]{FeCr$_2$S$_4$}

\begin{document}

\title{FeCr$_2$S$_4$ in magnetic fields: possible evidence for a multiferroic ground state}
\date{\today}

\author{J.~Bertinshaw}
\author{C.~Ulrich}
\thanks{Correponding author C.U. (c.ulrich@unsw.edu.au)}
\affiliation{The School of Physics, The University of New South Wales, Sydney, NSW 2052, Australia}
\affiliation{Australian Nuclear Science and Technology Organisation, Lucas Heights, NSW 2234, Australia}

\author{A.~G\"{u}nther}
\author{F.~Schrettle}
\author{M.~Wohlauer}
\author{S.~Krohns}

\affiliation{Experimental Physics~V, Center for Electronic
Correlations and Magnetism, University of Augsburg,
D-86135~Augsburg, Germany}

\author{M.~Reehuis}
\affiliation{Helmholtz-Zentrum Berlin f\"{u}r Materialien und Energie, D-14109 Berlin, Germany}

\author{A.~J.~Studer}
\author{M.~Avdeev}
\affiliation{Australian Nuclear Science and Technology Organisation, Lucas Heights, NSW 2234, Australia}

\author{D.~V.~Quach}
\affiliation{Department of Chemical Engineering and Materials
Science, University of California, Davis, CA 95616, USA}

\author{J.~R.~Groza}
\affiliation{Department of Chemical Engineering and Materials
Science, University of California, Davis, CA 95616, USA}

\author{V.~Tsurkan}
\affiliation{Experimental Physics~V, Center for Electronic
Correlations and Magnetism, University of Augsburg,
D-86135~Augsburg, Germany} \affiliation{Institute of Applied
Physics, Academy of Sciences of Moldova, MD-2028~Chisinau, Republic
of Moldova}

\author{A.~Loidl}
\author{J.~Deisenhofer}
\thanks{Correponding author J.D. (joachim.deisenhofer@physik.uni-augsburg.de)}
\affiliation{Experimental Physics~V, Center for Electronic
Correlations and Magnetism, University of Augsburg,
D-86135~Augsburg, Germany}

\maketitle

{\bf
We report on neutron diffraction, thermal expansion, magnetostriction, dielectric, and specific heat measurements on polycrystalline \fcs\ in external magnetic fields.
The ferrimagnetic ordering temperatures $T_{\mathrm{C}}\approx 170$\,K and the transition at $T_{\mathrm{OO}}\approx 10$\,K, which has been associated with orbital ordering, are only weakly shifted in magnetic fields up to 9\,T. The cubic lattice parameter is found to decrease when entering the state below $T_{\mathrm{OO}}$. The magnetic moments of the Cr- and Fe-ions are reduced from the spin-only values throughout the magnetically ordered regime, but approach the spin-only values for fields $>$5.5\,T. Thermal expansion in magnetic fields and magnetostriction experiments indicate a contraction of the sample below about 60\,K. Below $T_{\mathrm{OO}}$ this contraction is followed by a moderate expansion of the sample for fields larger than $\sim$4.5\,T. The transition at $T_{\mathrm{OO}}$ is accompanied by an anomaly in the dielectric constant. The dielectric constant depends on both the strength and orientation of the external magnetic field with respect to the applied electric field for $T<T_{\mathrm{OO}}$. A linear correlation of the magnetic-field-induced change of the dielectric constant and the magnetic-field dependent magnetization is observed. This behaviour is consistent with the existence of a ferroelectric polarization and a multiferroic ground state below 10\,K.}


\newpage

\section*{Introduction}

The interplay of orbital, lattice, and spin degrees of freedom is of fundamental importance in understanding the enormous variety of phenomena and ground states in transition-metal compounds.\cite{Kugel1982,Feiner1982b,Tokura2000} In particular, the effects of short and long-range orbital ordering or of dynamic and static cooperative Jahn-Teller (JT) distortions on the magnetic ground state have been in the focus of condensed-matter physics throughout the research activities on, for example, cuprates and manganites. The respective JT active magnetic ions Cu$^{2+}$ and Mn$^{3+}$  usually exhibit a strong JT coupling in octahedral environment and contribute to cooperative JT distortions and orbital ordering far above room temperature, e.g. in KCuF$_3$ and LaMnO$_3$.\cite{Kadota1967,Paolasini2002,Murakami1998} Similar electronic configurations are found in the case of compounds containing Fe$^{2+}$ or Cr$^{3+}$-ions in a tetrahedral crystal-field,\cite{Feiner1982b,Feiner1982a,Wang2011,Wulferding2011,Wang2012} but the electron-lattice coupling is generally weaker and competes with spin-orbit coupling and exchange interactions, leading to strong fluctuations and frustration effects.

\fcs\ is a sulfur spinel that has been reported previously to exhibit a complex interplay of orbital correlations and magnetism, including ferrimagnetic ordering of the Cr and Fe sublattices below $T_{\mathrm{C}}\approx 170\,\mathrm{K}$~\cite{Shirane1964,Tsurkan2010} and a transition at $T_{\mathrm{OO}}=$ 10~K, which has been assigned to originate from orbital ordering.\cite{Englman1970,Spender1972,Brossard1979,Feiner1982a,Eibschutz1967,Hoy1968,Kim2002,Fichtl2005,Tsurkan2010} In addition, an anomaly in the low-field magnetization at $T_{\mathrm{M}}$ = 60~K revealed a change in the magnetic configuration,\cite{Tsurkan2001a,Tsurkan2001b,Tsurkan2001c,Maurer2003,Mertinat2005,Shen2009,Tsurkan2010} which has recently been attributed to the formation of a non-collinear (possibly helical) spin configuration with an incommensurate modulation involving three different Fe sites as indicated by $\mu$SR, M\"{o}ssbauer, and time-resolved magneto-optical Kerr effect measurements.\cite{Kalvius2010,Engelke2011,Ogasawara2006} The same temperature has also been associated with the onset of short-range orbital order (orbital liquid) and dynamic JT distortions, suggesting a mutual influence of spin configuration and orbital correlations.\cite{Tsurkan2010}

At room temperature \fcs\ crystallizes in the spinel structure with space group $Fd\bar{3}m$ with eight formula units per unit cell. The Cr$^{3+}$-ions ($3d^3$, $S=3/2$) occupy the octahedral sites and the Fe$^{2+}$-ions ($3d^6$, $S\,=2$) are at the centre of S$^{2-}$ tetrahedron ions, forming a diamond lattice of two interlacing FCC sublattices.\cite{Goodenough1964} While the Cr sublattice is dominated by ferromagnetic exchange which is established through the $90^{\circ}$ Cr--S--Cr bond, the exchange coupling within the Fe sublattice is rather weak and overruled by a much stronger antiferromagnetic coupling to the Cr-ions, resulting in the ferrimagnetic arrangement. As the Cr$^{3+}$-ions with a half-filled $t_{2g}$ ground-state configuration do not possess orbital degrees of freedom, the orbital physics in this compound is dominated by the Fe$^{2+}$-ions. Because of the internal exchange field below $T_\mathrm{C}$, the five-fold spin degeneracy of the lower-lying $^5E$ states is lifted and the level splits into five orbital doublets. The lowest level is coupled to lattice vibrations, inducing a dynamic JT effect before an eventual formation of a static cooperative JT distortion of the sulfur tetrahedra below $T=10\,\mathrm{K}$.\cite{Goodenough1964,Gehring1975} Previous specific heat measurements show a $\lambda$-like anomaly at $T_{\mathrm{OO}}$, which was interpreted as evidence for an orbital order at low temperature.\cite{Tsurkan2010}
However, an unambiguous signature of a structural JT distortion, predicated by a deviation from cubic symmetry, remains evasive even in high-resolution structural x-ray studies. Thermal expansion measurements have revealed a contraction of the lattice at $T_{\mathrm{OO}}$.\cite{Tsurkan2010}

The influence of an external magnetic field on the magnetization of \fcs\ in the orbitally ordered phase was reported recently and it was proposed that there exists a new magnetic phase for fields larger than 5.5\,T, which might be related to the restoration of a commensurate collinear spin structure.\cite{Ito2011} Related compounds FeCr$_2$O$_4$ and CoCr$_2$O$_4$ (with orbital degrees of freedom) have been found to be multiferroic,\cite{Yamasaki2006,Singh2011} and CoCr$_2$O$_4$ exhibits unconventional properties as a function of applied magnetic field. Therefore, we focus in this study on the influence of external magnetic fields on the structural, magnetic, and dielectric properties of \fcs\ in the orbitally ordered phase. We identify a decrease of the cubic lattice parameter below $T_{\mathrm{OO}}$ and a step-like increase of the magnetic moments of the Cr- and Fe-ions upon application of an external magnetic field. Moreover, our dielectric measurements in magnetic fields provide evidence for the existence of a ferroelectric polarization in \fcs. All investigated quantities show changes in magnetic fields of about 5~T.

\section*{Results and Discussion}
\subsection*{Neutron Diffraction}

\begin{figure*}[t]
\centering
\includegraphics[keepaspectratio,width=17cm]{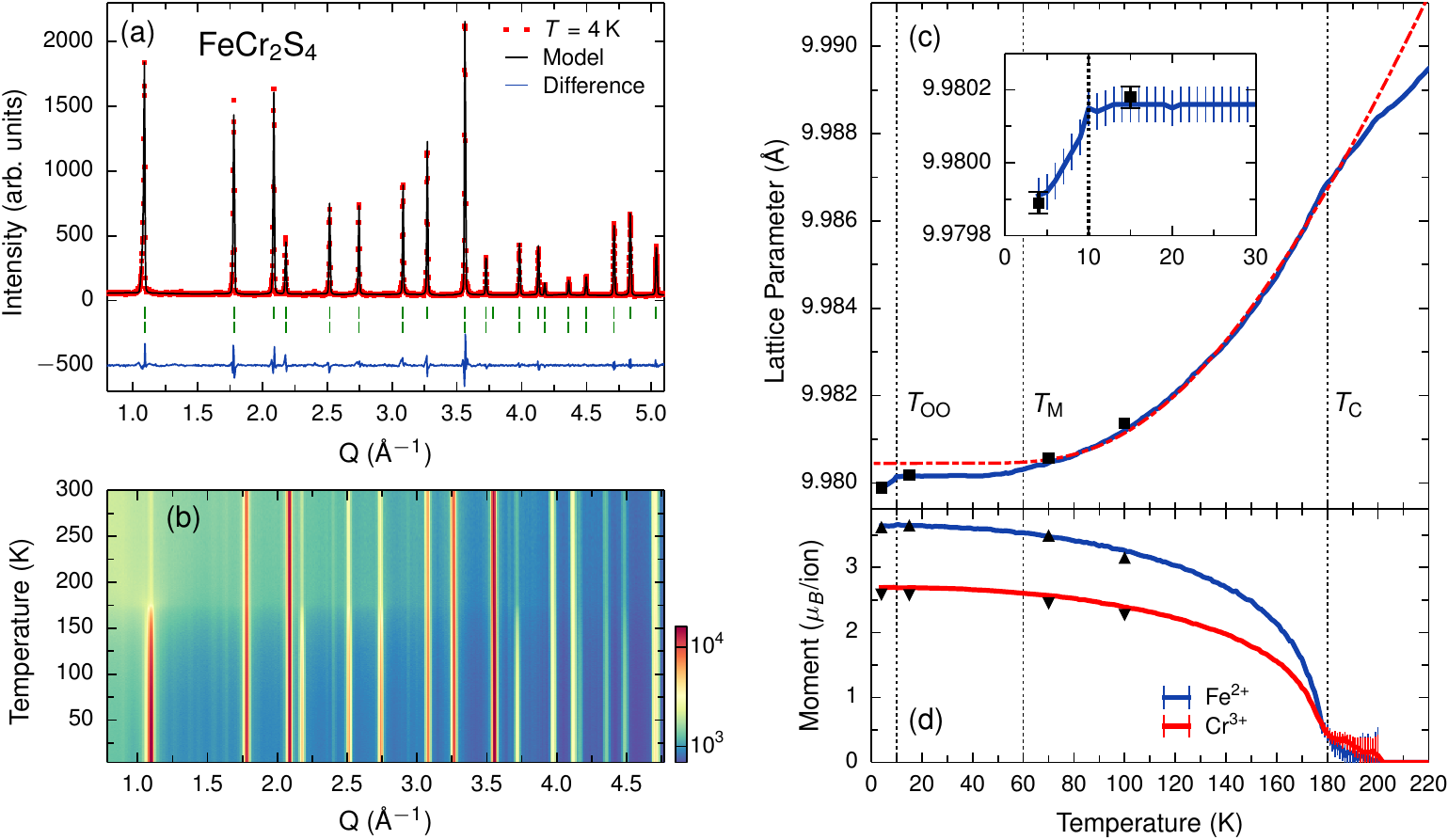}
\caption{Results of the neutron diffraction temperature dependency study. (a)  Rietveld refinement of high resolution data collected at 4\,K indicates a cubic spinel crystal structure and collinear antiparallel spin alignment. (b) High intensity neutron diffraction measured from $\mathrm{T}=4$\,K to 300\,K represented in a contour plot with a logscale color map (counts per 10 min) to enhance subtle modifications, revealing $T_{\mathrm{C}}\approx180\,$K. (c) Temperature dependency of the refined cubic lattice parameter. The dashed line corresponds to the behavior of a normal anharmonic solid, from which \fcs\ deviates due to short range orbital order below $T_{\mathrm{M}}\approx 60\,$K. Inset: Contraction of the lattice below $T_{\mathrm{OO}}=10\,$K.  (d) Temperature dependency of the magnetic moment parameter of the Fe- and Cr-ions. Above 200\,K the magnetic phase was not included in the refinement. In (c) and (d) the solid lines correspond to data collected on Wombat, and black markers correspond to high resolution data collected on the instrument Echinda.
}
\label{fig:nd}
\end{figure*}

\begin{table}[b]
\centering
\small
\caption{Results of the crystal structure refinements of \fcs\ from neutron powder diffraction ($\lambda = 2.4395$\,\AA). The structure refinements of the data sets collected at 4 and 15\,K were carried out in the cubic space group $Fd\bar{3}m$. The nuclear and magnetic residuals are defined as $R_{\mathrm{N}} = \sum{||F_{obs}| - |F_{cal}||}/\sum{|F_{obs}|}$ and $R_{\mathrm{M}} = \sum{||I_{obs}| - |I_{cal}||}/\sum{|I_{obs}|}$, respectively.}
\label{table:nd_ech_result}
\begin{tabular*}{0.48\textwidth}{@{\extracolsep{\fill}}lcc@{}}
\hline
Temperature           &4\,K       &15\,K    \\
\hline
$a = b = c$ [\AA]     &9.97989(3) &9.98018(3) \\
$\mu$(Fe) [$\mu_B$]   &--3.69(6)  &--3.72(7)  \\
$\mu$(Cr) [$\mu_B$]   &2.64(3)    &2.63(3)    \\
$x$(S)                &0.2585(3)  &0.2586(3)  \\
$R_{\mathrm{N}}$       &2.63       &2.30       \\
$R_{\mathrm{M}}$       &4.68       &4.39       \\
\hline
\end{tabular*}
\end{table}

\begin{table*}
\centering
\small
\caption{Comparison of the derived magnetic moments of the Fe- and Cr-ions and the total magnetic moment per formula unit (in units of the Bohr magneton $\mu_B$) from neutron diffraction (ND), x-ray spectroscopy (XPS), magnetization measurements (Mag.), and theoretical approaches (Calc.) for lowest temperatures.}
\label{table:mag_compare}
    \begin{tabular*}{0.95\textwidth}{@{\extracolsep{\fill}}lccccccccc@{}}
    \hline
    ~         & This Work & ND\cite{Kim2002} & ND\cite{Shirane1964} & XPS\cite{Kurmaev2000} & Mag.\cite{Zestrea2008}& Mag.\cite{Tsurkan2001b}& Mag.\cite{Ito2011} & Calc.\cite{Park1999}  & Calc.\cite{Sarkar2010}\\ \hline
    Fe$^{2+}$ & $-3.69$   &$-3.52$           &$-4.2$                &$-3.89$                & -                     & -                      & -                  &$-3.70$                &$-3.40$                \\
    Cr$^{3+}$ & 2.64      & 2.72             & 2.9                  & 2.71                  & -                     & -                      & -                  & 3.11                  & 2.66                  \\
    Total/f.u.& 1.59      & 1.92             & 1.6                  & 1.53                  & 1.52                  & 1.84                   & 1.6                &  2.0                   & 1.92                  \\
    \hline
    \end{tabular*}
\end{table*}

\begin{figure*}[t]
\includegraphics[keepaspectratio,width=17cm]{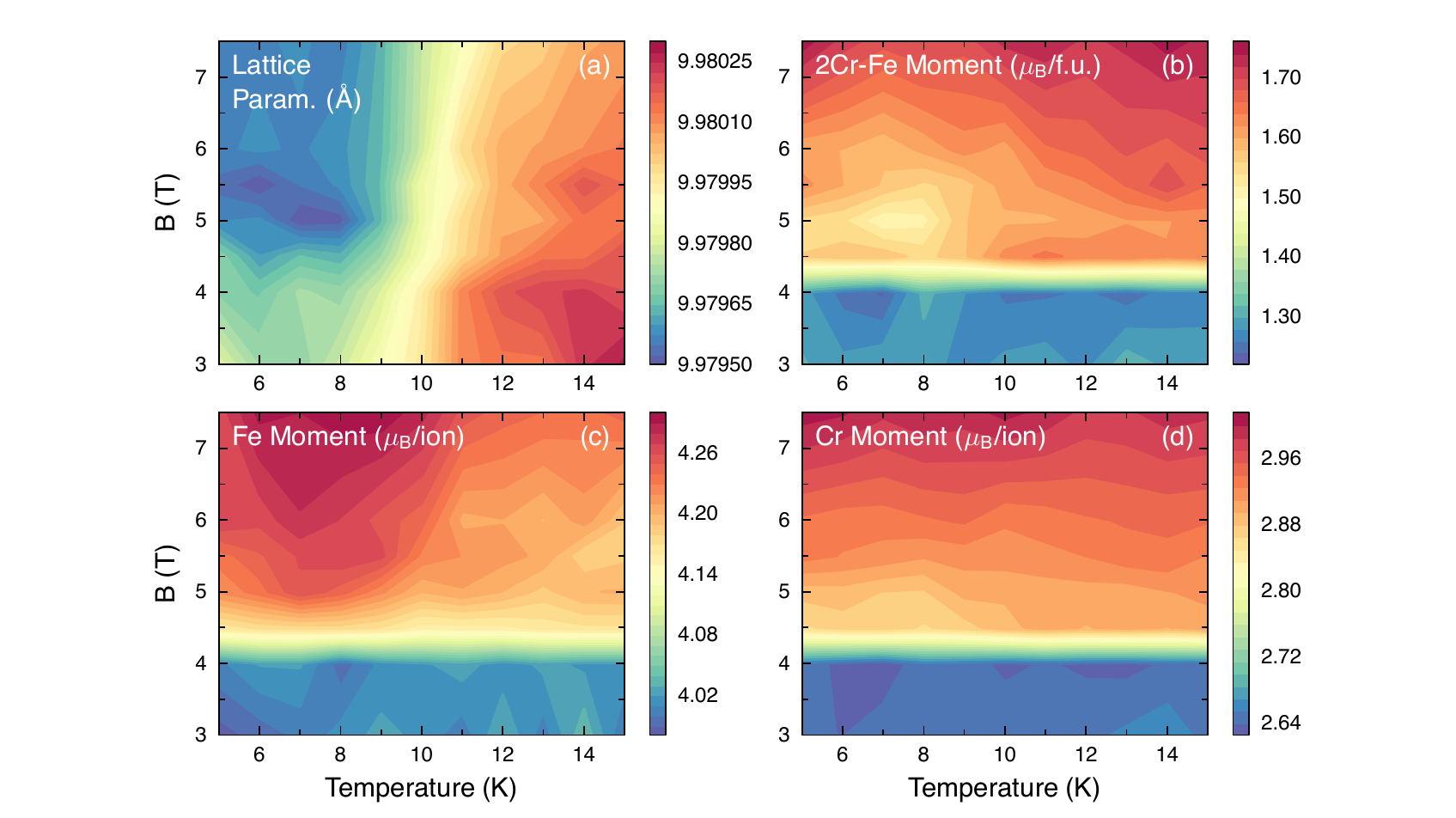}
\caption{Results of the magnetic field dependency study have been plotted as contour maps. (a) The cubic lattice parameter shows a clear contraction at T$=10\,$K for all magnetic fields studied. The refined (b) bulk magnetic moment, (c) Fe-ion moment and (d) Cr-ion moment parameters reveal a clear increase at around 4.5\,T, which is interpreted as the realization of full spin-only values under an applied magnetic field of the order of the anisotropy field, as described in the text.}
\label{fig:nd_wom_mdep}
\end{figure*}

High resolution neutron diffraction patterns were collected on the instrument Echidna, located at the Bragg Institute, ANSTO, at temperatures of $\mathrm{T}=4\,$K, 15\,K, 70\,K, and 100\,K. Figure~\ref{fig:nd}(a) presents raw data collected at 4K, together with the final Rietveld refinement results. Refinements of the diffraction patterns were performed in the high-symmetric cubic spinel structure with the space group $Fd\bar{3}m$ (no. 227). The atoms are located at the following Wyckoff positions: Fe at $8a$ ($\frac{1}{8}$, $\frac{1}{8}$, $\frac{1}{8}$), Cr at $16d$ ($\frac{1}{2}$, $\frac{1}{2}$, $\frac{1}{2}$), and S at $32e$ ($x, x, x$). Instrumental parameters were fitted using the diffraction pattern measured at 4\,K, and fixed for the refinement of the succeeding datasets. Fitted structural parameters of the Rietveld refinements for 4\,K and 15\,K are listed in Table \ref{table:nd_ech_result}. Within the resolution limits of the instrument the sulfur position parameter $x$ did not change between 4 and 15\,K, but there is a notable decrease of the cubic lattice parameter between $T=4\,$K and 15\,K as shown in the inset of Fig.~\ref{fig:nd}(c).

Therefore, we further investigated this trend in a detailed neutron diffraction study. Diffraction patterns were collected on the high intensity instrument WOMBAT, located at the Bragg Institute, ANSTO, at temperatures from 4\,K to 300\,K in 1\,K steps, which revealed the transition to ferrimagnetic order below $T_{\mathrm{C}}\approx180\,$K. The full 4\,K to 300\,K diffraction dataset was systematically refined to the cubic spinel space group $Fd\bar{3}m$. Initial crystal and magnetic phase parameters were set using the results of the refinement of the high-resolution data. Instrumental characteristics such as peak shape parameters were fitted for the 4\,K dataset and fixed for succeeding temperature refinements. Above 200\,K the magnetic phase was not included in the refinement. The refined cubic lattice parameter and magnetic moments of the Cr- and Fe-ions are plotted in Fig.~\ref{fig:nd}(c) and \ref{fig:nd}(d), respectively, which demonstrate the consistency between the temperature dependency of the high-intensity and high-resolution neutron diffraction studies. The insert of Fig.~\ref{fig:nd}(c) reveals a distinct reduction of the cubic lattice parameter below 10\,K. This contraction can be interpreted as a microscopic signature of the antiferro-orbital long-range order below $T_{\mathrm{OO}}=10\,$K. In Fig.~\ref{fig:nd}(c) we show that the temperature-dependent trend in the lattice parameter below $T_{\mathrm{C}}$ to $T_{\mathrm{M}}$ at 60\,K is well described by normal anharmonic lattice dynamics, and was fit accordingly to $V(\mathrm{T})=V_O[1+A/[\exp(\theta/\mathrm{T})-1]]$, where $V_O$ represents the cell volume extrapolated to $0\,\mathrm{K}$, $\theta$ is an averaged Debye temperature ($\theta = 481$\,K in the ferrimagnetic region), and $A$ is a fitting constant. In the transition from paramagnetic to ferrimagnetic order at $T_{\mathrm{C}}$ a deviation in the lattice parameter contraction is found to form as a result of magnetostriction. More intriguing is a deviation from the fitted curve for a normal anharmonic solid below 60\,K, which indicates enhanced shrinking of the lattice parameter. To our knowledge these deviations in structural characteristics have not been microscopically identified using neutron diffraction prior to this study. However, they are consistent with previous bulk techniques such as thermal expansion measurements.\cite{Tsurkan2010} The behavior below $T_{\mathrm{M}}\approx 60\,\mathrm{K}$ has previously been associated with short-range orbital fluctuations \cite{Tsurkan2010} and the emergence of a non-collinear spin configuration.\cite{Kalvius2010}

The temperature dependency of the magnetic moment of the Fe- and Cr-ions as determined from the refinement of the neutron diffraction data are presented in Fig.~\ref{fig:nd}(d). The magnetic phase symmetry was defined to align the Fe and Cr moments antiparallel along the structural $c$-direction as our data indicates that the spins are aligned along the cubic crystal axes. No additional Bragg reflections, which would indicate a weak spin canting or incommensurate structure, were detected within the experimental resolution. Simulations of a possible spin canting did indicate that additional transversal magnetic components of no larger than 0.3~$\mu_B$/Fe$^{2+}$-ion could exist below 60 K. The obtained values of $-3.69\,\mu_B$ and $2.64\,\mu_B$ for the Fe- and Cr-ions at 4\,K, respectively, are consistent with values recently determined in a neutron diffraction study,\cite{Kim2002} along with other experimental and theoretical approaches, which are compared in Table~\ref{table:mag_compare}. In particular, density-functional based calculations found reduced spin moments of $-3.3\,\mu_B$ and $2.7\,\mu_B$ for both types of ions and an orbital moment of $-0.13\,\mu_B$ on the Fe-ions. \cite{Sarkar2009,Sarkar2010} This is in agreement with our resulting total magnetic moments within experimental error. The orbital contribution of the Fe-ions is argued to originate from spin-orbit coupling,\cite{Sarkar2009} which is in agreement with the single-ion picture, where the second-order contribution of spin-orbit coupling yields an effective $g$-factor of $g=2.1$.\cite{Gibart1969} The successful refinement of the \fcs\ magnetic structure indicated that the spin-only moments are reduced from expected values of $-4.2\,\mu_B$ for Fe$^{2+}$ with $S=2$ and $g=2.1$ and $3.0\,\mu_B$ for Cr$^{3+}$ with spin $S=3/2$ and $g=2$. Moreover, recent reports have indicated modifications to the bulk magnetic properties of \fcs\ under large applied magnetic fields in this temperature range.\cite{Ito2011,Shen2011}

Neutron diffraction was performed on \fcs\ under external applied magnetic fields of up to 7.5\,T in order to investigate a potential link between orbital order, the magnetic and crystal structures and external fields. In particular, deviations in the magnetic and crystal structure were studied by performing Rietveld refinement of the lattice and magnetic moment parameters of the Fe$^{2+}$ and Cr$^{3+}$-ions. Neutron diffraction patterns were collected on the instrument Wombat at ANSTO under applied magnetic fields from 3\,T up to 7.5\,T in 0.5\,T steps at temperatures from 5\,K up to 15\,K in 1\,K steps. For the Rietveld refinements, the collected diffraction patterns were treated using the cubic spinel structure and collinear antiparallel magnetic structure as determined previously, since there was no evidence for additional Bragg peaks within the experimental resolution. Crystal and magnetic structural parameters were initially set based upon the results of the zero-field refinement results (see Table \ref{table:nd_ech_result}). Under an initial applied field of 0.5\,T, the Fe- and Cr-ion spins structure of cubic \fcs\ preferentially rotates to align along the crystal axis in the applied magnetic field direction. Additional parameters to account for this preferential orientation of the magnetic phase were implemented in the refinement, resulting in a significant improvement to the fitting residuals.

The obtained lattice parameter and magnetic moments are presented in Fig.~\ref{fig:nd_wom_mdep}. In Fig.~\ref{fig:nd_wom_mdep}(a) the reduction of the cubic lattice parameter below $T_{\mathrm{OO}}$ observed in zero-field remains visible under applied magnetic fields up to 7.5\,T. This indicates that the phase below $T_{\mathrm{OO}}$ remains stable with respect to even relatively large external magnetic fields. A clear increase in the obtained magnetic moment of the Fe$^{2+}$ and Cr$^{3+}$-ions occurs at around 4.5\,T (Fig.~\ref{fig:nd_wom_mdep}b,c,d), with spin-only values of --4.2$\mu_B$ and 3.0$\mu_B$ approached for fields above 5.5\,T, respectively.

The reduced magnetic moments observed in neutron diffraction in zero-field can be explained due to the existence of a transverse quasi-paramagnetic component of the magnetic moment of the Fe$^{2+}$-ions. According to previous publications, they are a consequence of the orbital moment of the Fe-ions which causes an magnetocrystalline anisotropy with an estimated magnitude of about 2-4~T in the magnetically ordered regime $T<T_{\mathrm{C}}$. \cite{Sarkar2009,Sarkar2010,Ogasawara2006} A possible scenario to understand the observed increase is that these transverse components are suppressed in the external magnetic field and the magnetic moments reach the full spin-only values of the longitudinal magnetization. The strength of the necessary magnetic field is in good agreement with reports of a change in slope of the magnetization as a function of field \cite{Ito2011,Shen2011} and with the order of magnitude of the field necessary to overcome the magnetocrystalline anisotropy in \fcs.\cite{Ogasawara2006} Therefore, we believe that for fields larger than 5.5~T FeCr$_2$S$_4$ becomes an isotropic ferrimagnetic. At the same magnetic field strength the magnetostriction and the dielectric properties also show distinct changes, as discussed in the following.

\subsection*{Thermal Expansion, Magnetostriction, and Specific Heat}

\begin{figure*}[t]
\includegraphics[keepaspectratio,width=17cm]{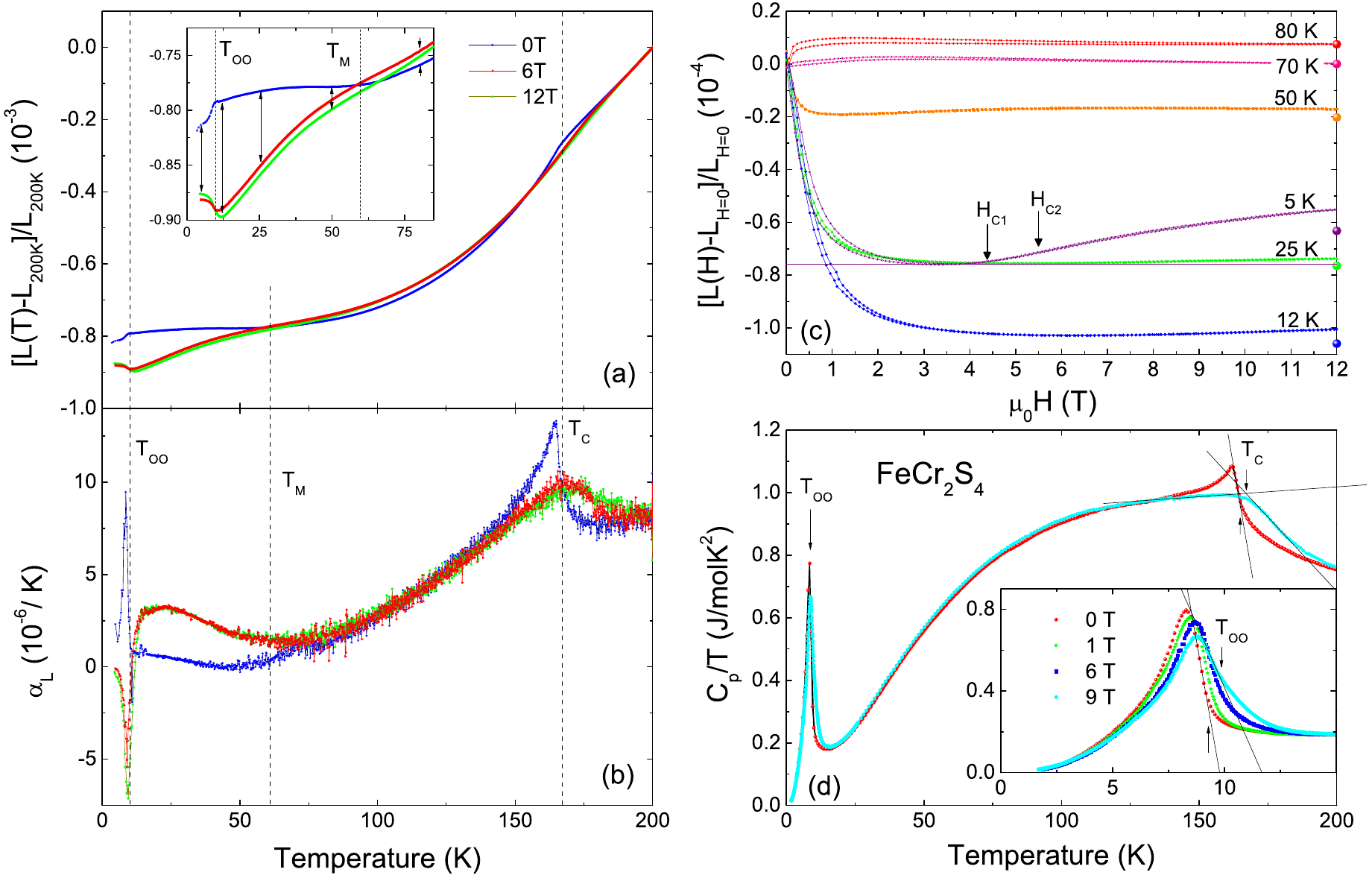}
\caption{Temperature dependency of (a) the thermal expansion normalized to the value at 200\,K and (b) the thermal expansion coefficient $\alpha$ in magnetic fields of 0, 6, and 12\,T. Inset in (a): thermal expansion for $T\leq 80$\,K on an enlarged scale. Arrows indicate selected temperatures where the differences in the thermal expansion in constant magnetic fields are compared to the thermal expansion vs. magnetic field normalized to the zero-field value measured at different temperatures shown in (c). The values derived from the temperature dependencies as described in the text are indicated by large spheres. The arrows at $H_\mathrm{c1}$ and $H_\mathrm{c2}$ mark the fields associated with the beginning expansion along the direction of the magnetic field. (d) Temperature dependency of the specific-heat in zero and 9\,T plotted as $C_P/T$~vs.~$T$ showing the ferrimagnetic and orbital ordering transitions at $T_{\mathrm{C}}$ and $T_{\mathrm{OO}}$, respectively. Inset: Changes of the anomaly at the orbital-ordering transition for 0, 1, 6, and 9\,T. Solid lines indicate linear fits to estimate the transition temperatures as described in the text.}
\label{fig:DL_alpha_C}
\end{figure*}

The temperature dependence of the thermal expansion $(L(T)-L_{\mathrm{200 K}})/L_{\mathrm{200 K}}$ of the sample's thickness $L$ in the direction of the applied magnetic fields of a dense polycrystalline sample obtained by spark-plasma sintering are shown in Fig.~\ref{fig:DL_alpha_C}(a) together with zero-field data reported in Ref.~\onlinecite{Tsurkan2010}. In the paramagnetic phase at 200\,K we do not expect any significant magneto-strictive effects enabling us to compare directly the effect of the external magnetic field, by normalizing the data to the thickness $L_{\mathrm{200 K}}$ of the sample at 200~K.  The temperature dependence in zero field clearly shows a contraction of the sample upon entering the state below 10\,K,\cite{Tsurkan2010} which is in agreement with the temperature evolution of the lattice parameter in Fig.~\ref{fig:nd}(c).

The effect of the external magnetic field above 100\,K is small, but a clear broadening of the anomaly at the ferrimagnetic transition is found in the linear thermal-expansion coefficient $\alpha_L=1/L(\partial L/\partial T)$ for the direction parallel to the applied magnetic field (Fig.~\ref{fig:DL_alpha_C}(b)). In the temperature range $60\leq T\leq 130$\,K only slight deviations from the zero-field curve can be detected. However, below about 60\,K the sample contracts much stronger when a magnetic field is applied. This temperature coincides with the temperature $T_{\mathrm{M}}$ where a transition from a collinear to a non-collinear spin configuration with three different Fe sites has been reported by $\mu$SR and where orbital fluctuations are assumed to set in.\cite{Kalvius2010,Tsurkan2010,Engelke2011}
The relative change of the thermal expansion upon entering the ground state below 10\,K is similar to the zero-field data, but under applied fields of 6\,T and 12\,T the sample expands again towards lowest temperatures. This is clearly seen in the negative thermal expansion coefficient for the transition in magnetic fields. The temperatures of the minima in the thermal expansion in magnetic fields are used as a measure of the orbital-ordering transition temperatures.

The relative magnetostriction effect of the sample  $(L(\mathrm{H})-L_{\mathrm{H}=0})/L_{\mathrm{H}=0})$ is shown in Fig.~\ref{fig:DL_alpha_C}(c) for selected temperatures, which are also indicated by arrows in the inset of Fig.~\ref{fig:DL_alpha_C}(a). The values at 12\,T are compared with the values (large solid symbols in Fig.~\ref{fig:DL_alpha_C}(c)) obtained by subtracting the corresponding temperature dependencies in Fig.~\ref{fig:DL_alpha_C}(a). The different data sets agree well. The deviations at lower temperatures might indicate small variations in temperature during the different measurements. At high temperatures the field effects are small and the almost field independent zero-magnetostriction curve at 70\,K indicates a sign change as expected from the temperature dependence of the thermal expansion. Accordingly, the largest contraction is observed at 12\,K, just above the orbital ordering transition. At both 12\,K and 25\,K the lattice contracts strongest in field from zero to about 2\,T, undergoes a very broad shallow minimum, and tends to saturate up to the highest measured fields. In the orbitally ordered state a broad but clear minimum develops at around 3-4~T, signaling the change from a contraction to an expansion of the sample in higher magnetic fields. The external fields $B_{c1}$ and $B_{c2}$ associated with this change are estimated by the values where the magnetostriction curve increases visibly from the minimal value (indicated as a constant solid line) and by the inflection points, respectively.

From the neutron scattering results in magnetic field, it is clear that the volume contraction of the lattice below $T_{\mathrm{OO}}$  is still present in magnetic fields up to 7.5\,T. The strong effects seen in the macroscopic thermal expansion measurement must therefore be related to the existence of magnetic domains in the investigated sample. In fact, the largest changes occur in fields smaller than 2\,T (see Fig.~\ref{fig:DL_alpha_C})(c) at temperatures below $T_\mathrm{M} = 60\,$K, where strong domain formation has been reported \cite{Tsurkan2001c} in agreement with the observation of small hysteretic effects. Within the magnetic field range of our study, the contraction of the sample seems to saturate above 3-4\,T and above $T_{\mathrm{OO}}$, indicating that the magnetic domain structure is stabilized. In the orbitally ordered state, however, an expansion of the sample sets in between $B_{\mathrm{c}1}$ and $B_{\mathrm{c}2}$. As mentioned above this field range coincides with the increase of the magnetic moments of the Fe- and Cr-ions and reflects the order of magnitude of the magnetocrystalline anisotropy.

In Fig.~\ref{fig:DL_alpha_C}(d) we compare the temperature dependence of the specific heat in the representation $C_P/T$~vs.~$T$ in zero and 9\,T fields. While the sharp anomaly at the ferrimagnetic transition at 165\,K in zero field is strongly broadened at 9\,T, the changes of the $\lambda$-like anomaly around 10\,K at the orbital-ordering transition (shown in the inset for several external magnetic fields) exhibits only a small shift and a transfer of associated entropy to higher temperatures with increasing magnetic field. This indicates that the orbital configuration of the ground state is rather stable with respect to the applied magnetic field, but critical fluctuations are favored to occur at higher temperatures already. A shift of the anomaly and an entropy transfer at 14\,T has been reported, \cite{Shen2011} which is similar to the one we observe at 9\,T. The temperatures of the orbital-ordering transition at 9.2, 9.5, 9.9, and 10.3\,K  for 0, 1, 6, and 9\,T, respectively,  were evaluated from the $C_P/T$ dependencies in terms of the temperatures where the data starts to deviate from a linear fit at temperatures above the maximum (see inset of Fig.~\ref{fig:DL_alpha_C}(d)). The ferrimagnetic transition temperature in zero-field is determined by the same procedure as $T_{\mathrm{C}}=165$\,K. At 9\,T the transition temperature has been evaluated as the intersection point of two linear extrapolations below and above the broad maximum (see Fig.~\ref{fig:DL_alpha_C}(d)).

\begin{figure*}[t]
\includegraphics[keepaspectratio,width=17cm]{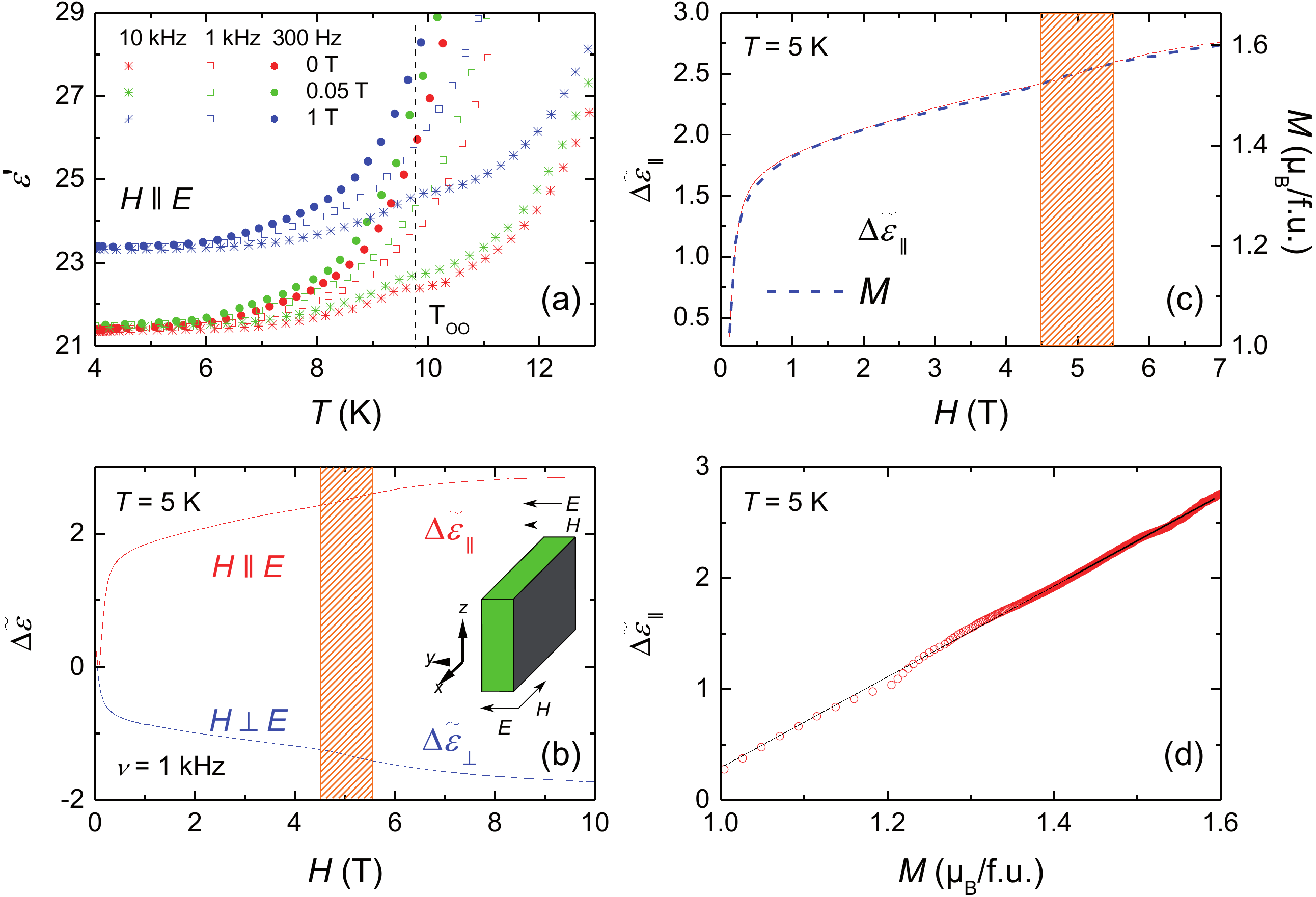}
\caption{(a) Temperature dependent dielectric constant $\epsilon'$ of polycrystalline \fcs~for selected frequencies. The measurement were performed in zero-magnetic field and with applied external magnetic fields $H$ (0.05~T and 1~T) parallel to the direction of the electric field $E$. (b) Magnetic field dependence of the polarization contribution $\Delta P(H)=P(H)-P(0)$ induced by the applied magnetic field (measured at $\nu = 1$\,kHz and $T = 5$\,K) for $H\parallel E$ and $H\perp E$. (c) Direct comparison of  $\Delta P$ vs. $H$ for $H\parallel E$ and magnetization $M$ vs. $H$. (d) Plot of $\Delta P(H)$ vs. $M(H)$ showing a linear correlation of the two quantities.}
\label{fig:eps_T_H}
\end{figure*}

\subsection*{Dielectric Spectroscopy}

In this section we present detailed measurements of the dielectric constant as a function of temperature and external magnetic fields specifically focusing on the occurrence of magnetoelectric effects. Temperature and frequency dependent of the dielectric loss has been reported earlier. In single crystals where the transition at 10\,K was suppressed evidence for a gradual freezing of the orbital degrees of freedom has been observed.\cite{Fichtl2005} However, in these crystals the conductivity was significantly enhanced compared to the samples investigated here and Maxwell-Wagner like relaxations extended down to the lowest temperatures.\cite{Lunkenheimer2002} Measurements of the dielectric constant as a function of temperature and magnetic field had not been reported.  Recently the occurrence of multiferroicity  in the orbitally ordered phase of FeCr$_2$S$_4$ has been discussed.\cite{Lin2013}
Figure~\ref{fig:eps_T_H}(a) shows the temperature dependence of the dielectric constant $\varepsilon^\prime$ of FeCr$_2$S$_4$ around the orbital ordering transition at $T_{\mathrm{OO}}\approx 10$\,K
for frequencies between 300 Hz and 10 kHz and external magnetic fields up to 1 T. In these measurements the magnetic field was oriented parallel to the electric field, $\mathbf{H}\parallel \mathbf{E}$. For all frequencies and magnetic fields the dielectric constant decreases on decreasing temperatures reaching constant low-temperature values below 6 K in the orbitally ordered state.

The orbital ordering transition in the 10 kHz-curve is indicated by a kink in the dielectric constant and is present for all applied magnetic fields. At lower frequencies the transition is masked by the increase of the dielectric constant towards higher temperatures due to an extrinsic Maxwell-Wagner relaxation.\cite{Lunkenheimer2002}  The strong increase of the dielectric constant in external magnetic fields of 1~T at low temperatures indicates a magnetic-field induced effect of the polarization and will discussed further in the following.

In Fig.~\ref{fig:eps_T_H}(b) we plot the change of the dielectric constant $\Delta \widetilde{\varepsilon}(H) =\varepsilon^\prime(H)-\varepsilon^\prime(0)$ induced by an applied magnetic field for the configurations $\mathbf{H}\parallel \mathbf{E}$ and $\mathbf{H}\perp \mathbf{E}$. The behavior is clearly nonlinear for both configurations. Notably, a positive contribution occurs for $\mathbf{H}\parallel \mathbf{E}$ and a smaller negative one for $\mathbf{H}\perp \mathbf{E}$. In both configurations we observe weak hysteretical behavior in low magnetic fields (not shown) and a change of slope in the applied magnetic field range $4.5 \mathrm{T}<H<5.5$~T (hatched area in Fig.~\ref{fig:eps_T_H}(b)) were observed. The latter effect is also documented in the neutron scattering and magnetostriction results discussed above. For $\mathbf{H}\parallel \mathbf{E}$ we plot $\Delta \widetilde{\varepsilon}(H)$ together with the magnetization $M(H)$ of the sample in Fig.~\ref{fig:eps_T_H}(c). Both curves can be scaled to fall on top of each other indicating $\Delta \widetilde{\varepsilon}(H) \propto M$, which even holds in the field range $4.5 \mathrm{T}<H<5.5$~T. This relation is confirmed in Fig.~\ref{fig:eps_T_H}(c), where $\Delta \widetilde{\varepsilon}(H)$ is plotted directly vs. $M(H)$ and reveals a linear correlation between the two quantities with a slope of 4.1 (f.u./$\mu_B$).

To understand this magnetic-field dependence we assume that the polarization of the sample as a function of electric and magnetic field may be written as
\begin{eqnarray}
\mathbf{P}(\mathbf{E},\mathbf{H})&=& \mathbf{P}^0(\mathbf{E},\mathbf{H}) + \varepsilon_0 \underline{\chi^e}\mathbf{E} +\underline{\alpha}\mathbf{H} +...
\end{eqnarray}
where $\mathbf{P}^0(\mathbf{E},\mathbf{H})$ is a possible spontaneous polarization, which may depend on the applied fields, $\underline{\chi^e}$ denotes the electric susceptibility tensor and $\underline{\alpha}$ the tensor of the linear magnetoelectric effect. Higher order terms in $\mathbf{E}$ and $\mathbf{H}$ are neglected. As the polarization exhibits a strongly nonlinear dependence on the applied field, we also neglect the linear magnetoelectric effect in the following.

In the dielectric measurement the dielectric tensor $\widetilde{\underline{\varepsilon_r}}$ is evaluated by assuming
\begin{eqnarray}
\mathbf{D}&=& \varepsilon_0 \mathbf{E} + \mathbf{P} = \varepsilon_0\widetilde{\underline{\varepsilon_r}} \mathbf{E}.
\end{eqnarray}
with $\mathbf{P}=\mathbf{P}^0(\mathbf{E},\mathbf{H})+ \varepsilon_0\underline{\chi^e} \mathbf{E}$. The presence of a static electric polarization $\mathbf{P}^0(\mathbf{E},\mathbf{H})$ may then result in an effectively magnetic-field dependent $\widetilde{\underline{\varepsilon_r}}$. Here, investigating a polycrystalline sample and $\mathbf{E}=(0,0,E_z)=E_z \mathbf{e_z}$ along the $z$-direction in the laboratory system we can distinguish two cases for the dielectric constant according to the relative orientation of the applied magnetic and electric fields:

\begin{eqnarray}
\Delta \widetilde{\varepsilon}_\parallel(H)&=& \varepsilon_0^{-1}(\mathbf{P}^0(E_z,H_z)-\mathbf{P}^0(E_z,0)\cdot\mathbf{e_z})/E_z\\
\Delta \widetilde{\varepsilon}_\perp(H)&=& \varepsilon_0^{-1}(\mathbf{P}^0(E_z,H_y)-\mathbf{P}^0(E_z,0)\cdot\mathbf{e_z})/E_z
\end{eqnarray}
The experiment shows that $\Delta \widetilde{\varepsilon}_\parallel(H)>0$ while $\Delta \widetilde{\varepsilon}_\perp(H)<0$, which originates from a change of the polarization component along the $z$-axis in the presence of the applied magnetic field. Applying the magnetic field perpendicular to the electric field obviously reduces the projection of the total polarization on the $z$-axis and leads to $\Delta \widetilde{\varepsilon}_\perp(H)<0$.

Given the linear relation between $\Delta \widetilde{\varepsilon}_\parallel(H)$ and the magnetization $M(H)$ we conclude that within our assumptions the $z$ component  $\mathbf{P}^0(E_z,H_z)\cdot\mathbf{e_z}$ of the polarization and the corresponding ferroelectric domains undergo the same changes as the magnetic domains and that ferroelectric and ferrimagnetic domains are linked to each other in \fcs.

A similar situation has recently been reported in single crystals of the  multiferroic spinel CoCr$_2$O$_4$, \cite{Yamasaki2006,Choi2009} which is comparable to \fcs: it is isostructural at room temperature, becomes a collinear ferrimagnet below about 95~K and ferroelectric below 27~K as a result of a conical-spiral magnetic ordering.\cite{Yamasaki2006,Choi2009} At the latter phase transition the dielectric constant in CoCr$_2$O$_4$ reportedly exhibits a small cusp similar to the effect shown in Fig.~\ref{fig:eps_T_H}(a). For CoCr$_2$O$_4$ a spin-driven origin has been proposed as the origin of multiferroicity.

Although the observed field dependence of the dielectric constant is not a direct proof of the existence of a static polarization component and multiferroicity, we believe that in FeCr$_2$S$_4$ a similar mechanism might be active. In our opinion, the clearly linear correlation between $\Delta \widetilde{\varepsilon}(H)\propto M(H)$ shown in Fig.~\ref{fig:eps_T_H}(c) strongly points towards a multiferroic ground state of FeCr$_2$S$_4$. However, the actual spin arrangement of FeCr$_2$S$_4$ below 50~K remains unsolved. Assuming a scenario where ferroelectric and ferrimagnetic domains are clamped, the field dependence of $\Delta \widetilde{\varepsilon}_\parallel(H)$ in Fig.~\ref{fig:eps_T_H}(c) in fields below 1~T can be understood in terms of ferroelectric domain reorientation and alignment effects, which are induced by the alignment of magnetic domains. The change of slope at $4.5 \mathrm{T}<H<5.5$~T is then related to the changes in the magnetization and the increase of the magnetic moments of the Fe- and Cr-ions assigned to the overcoming of the magnetocrystalline anisotropy.

\section*{Summary and Conclusions}

Using neutron diffraction we observed a reduction of the cubic lattice parameter below $T_{\mathrm{OO}}\approx 10$\,K in agreement with bulk measurements in thermal expansion. This transition was previously attributed to an onset of orbital order \cite{Tsurkan2010} and is not altered under an external magnetic field as demonstrated by our neutron diffraction and specific heat measurements. In the field range $4.5 \mathrm{T}<H<5.5$~T an increase of the magnetic moments of the Fe- and Cr-ions appears. The bulk magnetization accordingly exhibits a change of slope, and the magnetostriction causes an expansion of the sample along the direction of the applied field. These changes are attributed to the magnetocrystalline anisotropy, which is overruled by the external magnetic field, inducing an isotropic ferrimagnetic state for fields larger than 5.5~T. In this isotropic regime, the magnetic moments of the Fe and Cr reach the expected spin-only values, suggesting that transverse components of the magnetization are suppressed together with the magnetocrystalline anisotropy.

The dielectric measurements revealed a strong dependence on the applied magnetic fields. In case of parallel applied electric and magnetic fields, a linear correlation between the dielectric constant and the magnetization measured in magnetic field is established. We interpret this behavior as an indication for the existence of a ferroelectric polarization in FeCr$_2$S$_4$, where ferroelectric and magnetic domains are strongly coupled to each other. To corroborate such an interpretation we suggest that future studies may be directed to search for structural features of a polar symmetry group in \fcs.

\begin{figure}[t]
\centering
\includegraphics[keepaspectratio,width=8.9cm]{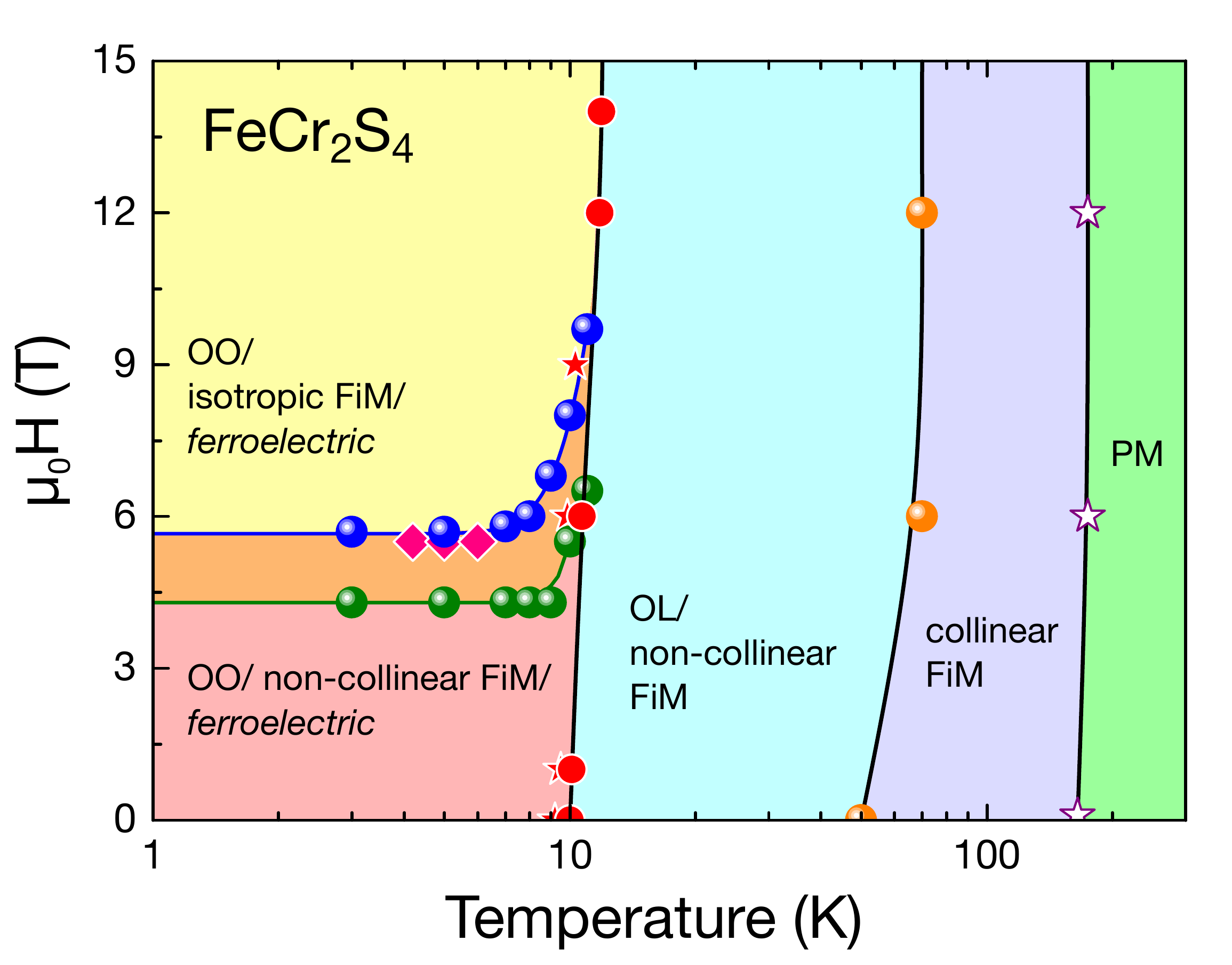}
\caption{$H-T$ phase diagram of \fcs\ (logarithmic temperature scale). OO: orbital order (as determined in Ref.~\onlinecite{Tsurkan2010}), FiM: ferrimagnet, OL: orbital liquid, PM: paramagnet. Values were taken from the respective temperature and field dependence of the thermal expansion and magnetostriction $\Delta L(H,T)$ (spheres) and the specific heat $C_P(T,H)$ (stars) and are in accordance with our neutron diffraction data presented in Fig. 2. Values from the magnetization $M(T,H)$ (diamonds) were taken from Ref.~\onlinecite{Ito2011}. Lines are to guide the eyes.}
\label{fig:phadia}
\end{figure}

Based on these results, we show a modified $H-T$ phase diagram of \fcs\ in Fig.~\ref{fig:phadia}. In the orbitally ordered state below about 10~K, we assign the observed features to a ferroelectric ground state and, hence, regarded FeCr$_2$S$_4$ as a multiferroic material like the related spinel CoCr$_2$O$_4$.\cite{Yamasaki2006,Choi2009} Most multiferroic materials are antiferromagnetic and the spin order cannot be addressed directly by an external magnetic field. This problem would be largely overcome by the predominantly ferrimagnetic structure of \fcs\ and would hence open new routes for basic research and applications in this potentially new multiferroic material.

The changes in the field range $4.5 \mathrm{T}<H<5.5$~T and for the lowest temperatures are assigned to a transition into an isotropic ferrimagnetic state. The corresponding necessary magnetic fields increase when approaching the orbital ordering temperature from below and reach up to about 10~T. Due to the increasing conductivity with temperature, no such clear anomalies were observed in the orbital liquid regime above 10\,K, but it can not be excluded that ferroelectric domains, which are only weakly clamped to the magnetic domains, are already present in the orbital liquid regime.

\section*{Methods}
Polycrystalline samples of \fcs\ for neutron experiments were prepared by solid-state reaction using high-purity elements. Post growth annealing under a chalcogen atmosphere was performed to achieve a pure stoichiometric ratio, which has been shown previously to exhibit the strongest low temperature anomaly.\cite{Tsurkan2010} For preparation of dense samples for thermal expansion and specific heat measurement, we used a spark-plasma-sintering (SPS) technique as described in Ref.~\onlinecite{Tsurkan2010}. While both samples exhibit the orbital ordering transition at about 10\,K, the sample obtained by SPS showed a reduced ferrimagnetic transition temperature of 165\,K in comparison to the one used for the neutron diffraction study with $T_{\mathrm{C}} = 180\,$K. Thermal-expansion, magnetostriction and dielectric experiments were carried out in a home-built setup using a capacitance bridge (AH 2700 Hagerling) for temperatures from 1.5-300\,K and in external magnetic fields up to 14\,T. The sample had the shape of a platelet and the expansion was measured along the direction of the applied magnetic field, which is perpendicular to the platelet. Dielectric measurements were perfomed using silverpaint contacts applied on both sides of the platelike sample, the magnetic field was applied perpendicular and parallel to the electric field. The heat capacity was measured in a physical properties measurements system (PPMS, Quantum Design) for temperatures between 1.8\,K and room temperature and in external magnetic fields up to 9\,T.

High resolution powder neutron diffraction measurements were performed using the instrument Echidna, located at the Bragg Institute, ANSTO, Sydney, Australia. Echidna is equipped with 128 $^3$He linear position sensitive detectors that are scanned in position to produce high resolution diffraction patterns with an angular step of $0.05^{\circ}$. A Ge monochromator was aligned on the $(3 3 1)$ reflection with a take-off angle of $140^{\circ}$ to select a wavelength of 2.4395\,\AA\ with a calculated accessible Q-range of 0.2-5.1\,\AA$^{-1}$. Collimation ensured a minimum FWHM resolution of $\sim$$0.4^{\circ}$. To achieve a good signal to noise ratio, each diffraction pattern was collected for 6 hours. The symmetry of the FeCr$_2$S$_4$ crystal and magnetic structures were characterized by performing a least-squares Rietveld refinement of the powder neutron diffraction data, using the {\sc fullprof} software suite.\cite{Rodriguez1993}

The high-intensity powder neutron diffractometer Wombat, located at the Bragg Institute, was used to conduct temperature and applied magnetic field dependency studies of the magnetic and crystal structures of \fcs. Magnetic fields up to 7.5\,T were applied using a vertical closed-cycle cryomagnet (Cryogenic Ltd.). Wombat uses a monolithic position-sensitive $^3$He detector, enabling fast collection times with little penalty to resolution. A vertically focused Ge monochromator was set to align on the $(1 1 3)$ reflection, with a $90^{\circ}$ take-off angle, giving an incident wavelength of 2.41\,\AA\ and an accessible Q-range of 0.5-4.9~\AA$^{-1}$. No collimation was used to maximize the count rate, resulting in a required 10 minute collection time per scan.

\begin{acknowledgments}
We thank K.-H. H\"{o}ck, I. Kezsmarki, H.-A. Krug von Nidda, B. Lake, M. Schmidt, and Z. Wang for fruitful discussions and A. Schmidt for technical assistance. We acknowledge partial support by the Deutsche Forschungsgemeinschaft via TRR 80 (Augsburg-Munich). M.~W. and S.~K. acknowledge the support by the Federal Ministry of Education and Research via project 03EK3015. J.~B. and C.~U. acknowledge the support of the Australian Institute of Nuclear Science and Engineering (AINSE) and the support of the Australian Research Council through the Discovery Projects funding scheme (project DP110105346).
\end{acknowledgments}

\section*{Author Contribution}

J.~B., C.~U., A.~G., M.~R. A.~J.~S., and M.~A. participated in the neutron diffraction experiment and preformed the data analysis.
V.~T., D.~V.~Q., and J.~R.~G. prepared and characterized the samples.
A.~G., F.~S., V.~T., and J.~D. conducted and analyzed specific heat, thermal expansion, and magnetization measurements.
M.~W. and S.~K. performed and analyzed the dielectric measurements.
All authors contributed to the interpretation of the data and to the writing of the manuscript.
C.~U., A.~L. and J.~D. conceived and supervised the project.

\section*{Additional Information}

The authors declare no competing financial interests.
Correspondence and requests for materials should be addressed to J.~B., C.~U., or J.~D.

\end{document}